# Electron transport in the single-layer semiconductor

Lianhua Zhang,[a] Ji'an Chen,[b] Fei Liu,[b] Zhengyang Du,[b] Yilun Jiang,[b] Min Han[b]*

Email：sjhanmin@nju.edu.cn

[a]*National Laboratory of Solid State Microstructures, School of Physics, and Collaborative Innovation Centre of Advanced Microstructures, Nanjing University, 210093 Nanjing, China.*

[b]*National Laboratory of Solid State Microstructures, College of Engineering and Applied Sciences, and Collaborative Innovation Centre of Advanced Microstructures, Nanjing University, 210093 Nanjing, China.*

**Abstract**

Two-dimensional (2D) materials are a new class of materials with interesting physical properties and applications ranging from nanoelectronics to sensing and photonics. In addition to graphene, the most studied 2D material, monolayers of other layered materials such as semiconducting dichalcogenides $MoS_2$ or $WSe_2$ are gaining in importance as promising channel materials for field-effect transistors (FETs) and phototransistors. However, it is unclear that how the specific process of electron transport is affected by temperature. So, nowadays the electron dynamics of single-layer semiconductor cannot be understood fundamentally. Here, we develop an analytical theory distinguishing from traditional energy band theory, backed up by Monte-Carlo simulations, that predicts the process of electron transport and the effect of temperature on the electron transport in the single-layer semiconductor. In this paper, A new model is built to deal with electron transporting in the sing-layer semiconductor. The resistance is decided by the barrier rather than the electron scattering in the single-layer semiconductor, which is macroscopic quantum effect. Electron transport in FETs with different dielectric configurations are investigated at different temperatures and a new control factor that is decided by top-gate voltage or bottom-gate voltage is introduced to describe the effect of gate voltage on the electron transport in 2D semiconductor. The results of simulation show the drain current is mainly determined by some elements, such as temperature, top-gate voltage, bottom-gate voltage and source-drain voltage. For FETs, the $I_{ds}$-$V_{ds}$ curves are non-linear and highly symmetric. However, the $I_{ds}$-$V_{bg}$ curves and the $I_{ds}$-$V_{tg}$ curves are also non-linear but highly asymmetric, which is very consistence with the experimental results. The transition mechanism from the insulator phase to metallic phase has been explained. For phototransistors, its light-induced electric properties are investigated in detail. Photocurrent generated from the phototransistor is solely determined by the illuminated optical power at a constant drain or gate voltage. The results show that photocurrent is increasing with the illuminated optical power and the $I_{ds}$-$V_{ds}$ curves are non-linear. For FETs and phototransistors, the number of exciton is increasing with temperature and the source-drain voltage. So maybe the field-effect mobility is decreasing at high temperature, which is related to the concentration of exciton. The exciton plays an important role in the single-layer semiconductor.

# 1. Introduction

Recently, 2D materials have attracted increasing attention. In particular, great interest has been focusing on the single-layer semiconducting materials, which exhibits the unique physical, optical and electrical properties correlated with its 2D ultrathin atomic layer structure [1-4]. Transition metal dichalcogenides (TMDs), a family of two-dimensional (2D) layered materials like graphene, have been a subject of tremendous amounts of experimental and theoretical studies due to their exciting electronic and optoelectronic properties [5-13]. Unlike thin fully-depleted silicon channels, physically limited by the oxide interface, single layer metal dichalcogenides are intrinsically 2D and, therefore, have no surface dangling bonds[14]. The monolayer thickness is constant, and the scale of the variations of the electrostatic potential profile perpendicular to the plane is only limited by the extent of the electronic wavefunctions. Hence, TMDs can in principle be considered immune to channel thickness modulation close to the drain. Building on these fundamental advantages, numerous field-effect transistor (FET) designs and phototransistors employing $MoS_2$ or $WS_2$ channels have been proposed. However, at present the success of TMD in electronics is limited by the difficulty in achieving high electronic mobilities. Nowadays a systematic study about the electronic transport mechanism of 2D semiconductor has not been done.

In this paper, we develop an analytical theory backed up the Monte-Carlo simulations, that predict the process of electron transport in 2D semiconductor. The effect of temperature on the characteristics of $I_{ds}$-$V_{ds}$ was also considered in our model. The dynamic behavior along the conduction paths can be analyzed quantitatively. In our theory the single-electron tunneling dynamics and the Coulomb attraction effect are used to investigate the electronic transport mechanism. These phenomenons can be explained by our theory, such as photoluminescence, photocurrent and electroluminescence.

## 2. The new model of electron transport

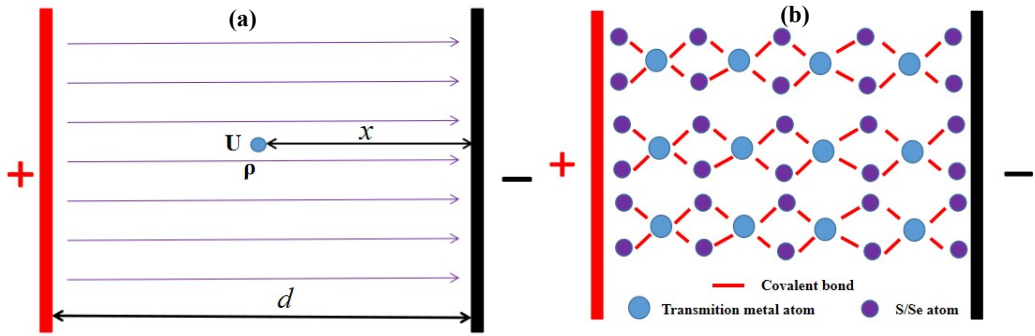

Fig. 1.(a) Parallel plate electrode system; (b) The schematic of 2D semiconductor.

Fig. 1a shows the parallel plate electrode system. In this system the relationship between the electric potential and the space charge density is following the Poisson equation:

$$\frac{d^2 U}{dx^2} = -\frac{\rho}{\varepsilon_0} \quad (1)$$

$\rho$ is the space charge density; $U$ is the electric potential.

In some point of the parallel plate electrode system, the relationship between the velocity of electron and the electric potential can be described:

$$\vec{v} = \sqrt{\frac{2eU}{m}} \vec{I} \quad (2)$$

$\vec{I}$ is the unit vector. So the current density is

$$\vec{j} = -\rho \vec{v} = -\rho \sqrt{\frac{2eU}{m}} \vec{I} \quad (3)$$

The space charge density is

$$\rho = -\vec{j} \sqrt{\frac{m}{2eU}} \vec{I} \quad (4)$$

So the Poisson equation can be described:

$$\frac{d^2U}{dx^2} = -\frac{\vec{j}}{\varepsilon_0}\sqrt{\frac{m}{2eU}}\vec{I} \tag{5}$$

In this system the electric potential of the cathode is following $U = 0$ and $\frac{dU}{dx} = 0$, the current density can be expressed after carrying out two integrations.

$$\vec{j} = \frac{4\varepsilon_0}{9}\sqrt{\frac{2e}{m}}\frac{U^{3/2}}{x^2}\vec{I} \tag{6}$$

At some point of the parallel plate electrode system, $x = d_a$, $U = U_a$

$$\vec{j}_a = \frac{4\varepsilon_0}{9}\sqrt{\frac{2e}{m}}\frac{U_a^{3/2}}{d_a^2}\vec{I} \tag{7}$$

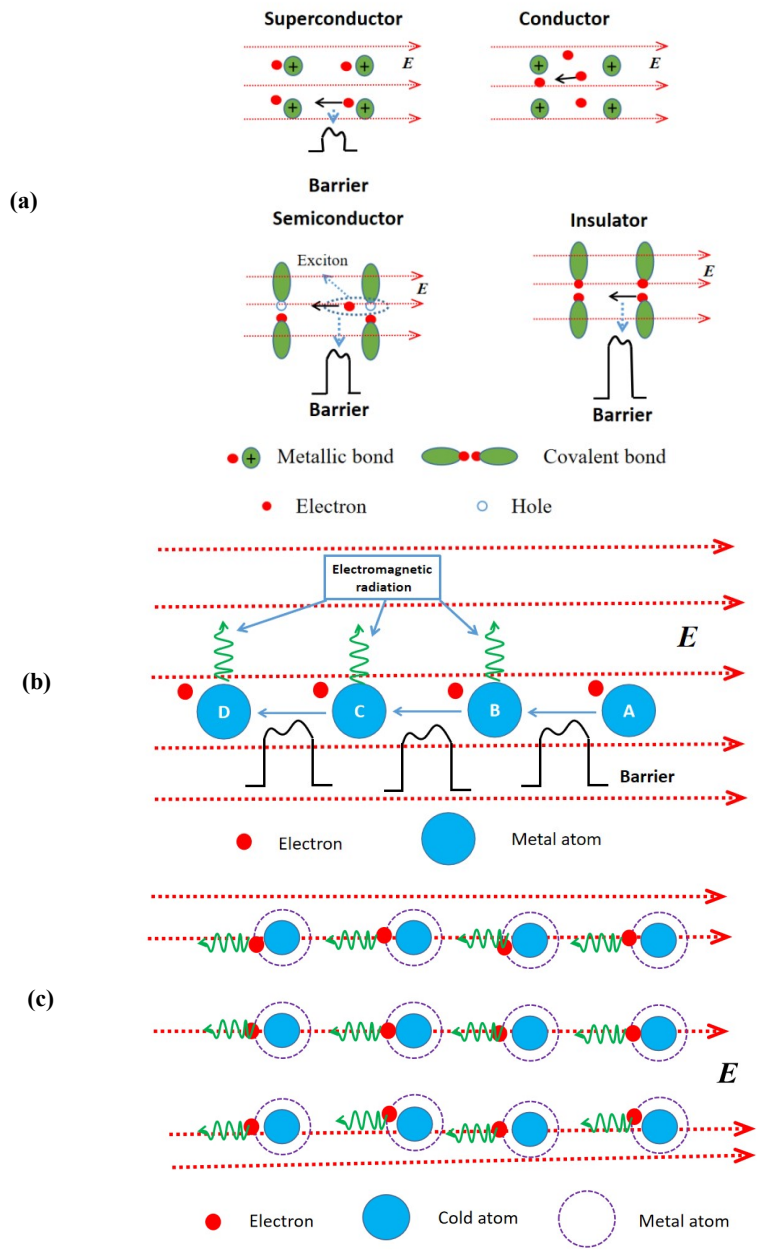

Fig. 2 (a) The new model of lattice: superconductor (left), conductor, semiconductor and insulator. (b) The process of electron entering into the atom with the emitted light. (c) the mechanism of superconductivity in the metal.

In this episode, we will use a new theory that is distinguishing from traditional energy band theory to investigate the mechanism of electron transport in four types of object: superconductor, conductor, semiconductor and insulator.

The electric current is formed by the regular directional motion of nearly-free electron in the conductor under the electric field in the classical model. When the electrons are moving in the conductor, the electrons are colliding with atomic core, which is the origin of resistance. The electron scattering can be solved by the Sommerfeld-Bloch individual-particle model and the electrons in the covalent bond are moving in the constant potential field. The resistance is the collective effect of electron scattering. However, in our theory when the electrons travel through a semiconductor, the interactive force between electron and atomic core is very strong and the height of the barrier is very high (see Fig. 2a). Its fundamental form of expression is covalent bond. Therefore the electrons have low probability to go over the barrier and can difficultly transport from one atomic core to the neighboring atomic core in the semiconductor and the excess energy is emitted by electromagnetic radiation (see Fig. 2b). Furthermore, temperature makes atomic cores dislocate, which will influence the current greatly. So the resistance is the collective effect of electron-hole pairs, which is macroscopic quantum effect. Temperature influences greatly the vibration of crystal lattice. When temperature is very high, the state of electronic transport can be changed from the semiconducting phase to the metal phase. However, the electrons can consciously transport in the path with the highest probability from one atomic core to the neighboring atomic core (see Fig. 2a). Lately we will demonstrate this conclusion. When the temperature drops to the absolute zero, based on the Fermi-Dirac distribution, all the electrons in the atom are located under the Fermi energy. The atomic radius will drop to the Fermi surface. The distance between atoms will become larger. Then it will have attraction force between atomic core and electron and the barrier will present. There is only an interaction between atomic core and electron. Based on the Pauli exclusion principle, two neighbouring electrons would have opposite spin orientation. Electron will move from one atomic core to the neighbouring atomic core. In this process the electrons will no energy-loss. This is origin of superconductivity (see Fig. 2c).

When the electrons travel through a semiconductor, the height of the barrier is relatively high and electrons are constrained. Its fundamental form of expression is covalent bond. The electrons have certain probability to go over the barrier and can transport in the semiconductor. For the insulator, electrons are tightly bound and the height of the barrier is the highest. So the electrons have little probability to go through the barrier. Therefore, it is very difficult to transport for electron in the insulator (see Fig. 2a). When temperature is improving, due to the weak bondage of the metallic bond, atoms in the conductor could easily happen the displacement unless the mass of atom is very heavy. The displacement plays an important role in the metallic phase. The displacement is very large when the temperature is high, So the current is decreasing with temperature. We will demonstrate this result lately. However, for the semiconductor, it is difficult to happen the displacement, which is owing to the strong bondage of the covalent bond. In high temperature or high bias voltage, the status of electron transport may be found the transition from the insulator phase to the metallic phase due to the great vibration of atomic core. So for some extreme conditions we can predict the transition could happen from insulator to conductor by our new theory.

When dealing with single-layer semiconductor between two outer electrodes (see Fig. 1b), the number of electron in the single-layer semiconductor is always integer and will change if a certain voltage is applied. Since only electrons can enter or leave the covalent bond as an entity, their number is also integer. Thus, electron transport in 2D semiconductor can be controlled by the external voltage (see Fig. 1b). If a covalent bond is charged by only one hole, its electronic potential will rise to attract charging. This process is characterized as "single electron tunneling" — Coulomb attraction phenomena in single-layer semiconductor (see Fig. 3).

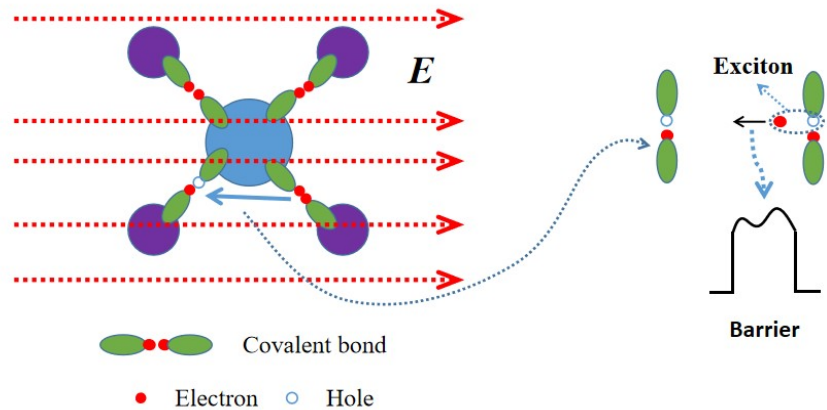

Fig. 3 The new model of electron transport in 2D semiconductor. The electron is transporting from one covalent bond to the neighboring bond with one hole.

In our model each covalent bond is independent, so it can be treated as a discrete phase model that an electron in some covalent bond passing through a barrier (see Fig. 3). Electron in the covalent bond is moving in the constant potential field. When the electron of one covalent bond is leaving from this covalent bond, an attractive force is generated to prevent the electron escaping from this covalent bond. This force can only work in a very short distance and the barrier is produced by this force. If an appropriate electric field is applied, the electron will have positive energy, then the electron has a certain chance to leave from this covalent bond. Exciton could happen in this process. So the barrier of exciton is dominated by the attraction force when electron is leaving from the covalent bond. So the transmission probability without an applied voltage is

$$P \sim e^{-2\sqrt{2mE_g}L/\hbar} \tag{8}$$

$E_g$ is energy gap, $L$ is the barrier width, $m$ is the effective mass of the electron and

$$E_g = V - E_a \tag{9}$$

In equation (9) $V$ is the barrier height and $E_a$ is the kinetic energy of the electron.

In 2D semiconductor each covalent bond is affected by single-electron tunneling dynamics and the Coulomb attraction effect. So when the electric potential of the covalent bond is $U_a$, the probability density of single electron transporting from this covalent bond to the neighboring covalent bond with one hole is

$$P \sim \frac{U_a^{3/2}\vec{I}}{d_a^2} e^{-2\sqrt{2mE_g}L/\hbar} \tag{10}$$

When the applied voltage between two electrodes is $U$, the uniform electric filed $E$ is present

$$E = U/d \tag{11}$$

where $d$ is the distance between two electrodes. The electric potential $U_a$ of some covalent bond is

$$U_a = Ud_a/d \tag{12}$$

$d_a$ is the distance between some covalent bond and cathode. So the transmission probability with applied voltage is

$$P \sim \frac{(U/d)^{3/2}\vec{I}}{d_a^{1/2}} e^{-2\sqrt{2mE_g}L/\hbar} \tag{13}$$

Under the condition of single-electron tunneling and the Coulomb attraction effect, the transmission probability is

$$P \sim \frac{(U/d)^{3/2}\cos\theta}{d_a^{1/2}}e^{-2\sqrt{2mE_g}L/\hbar} \tag{14}$$

$\theta$ is the relative angle between electric field line and the connecting line of two adjacent covalent bond, where electron could transmit from one covalent bond to the neighboring covalent bond with one hole. When the electron is entering into the neighboring covalent bond or returning to the old covalent bond, Excess energy is emitting by light, which is the underlying emission mechanism of electroluminescence (see Fig. 2b). All the photon should have same frequency. It is due to the observed photoluminescence and electroluminescence arising from the same excited state. It is the process of electrically excited luminescence in 2D semiconductor.

3. **The vibration model of lattice**

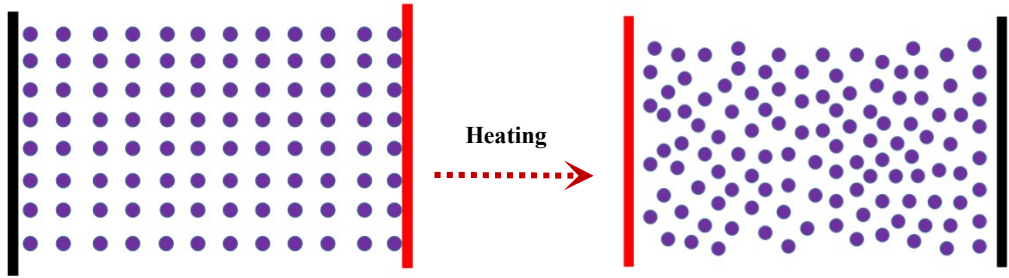

FIG. 4 The variation of lattice constant induced by temperature.

In this section we will investigate the effect of the lattice vibration on the electron transport. The displacement of atom was caused by the temperature. So the temperature is an important factor for the electron transport, therefore the influence of temperature should be taken into account in the simulation. Under the high temperature, not only atoms have enough energy to cause the dislocation and the degree of dislocation is related to the mass of atom in the metal. For the semiconductor, due to the strong covalent bond, the dislocation cannot easily happen but electrons in the covalent bond could have enough thermal kinetic energies to pass through the barrier and form the electron-hole pairs. After this process, the exciton could appear and the electroluminescence could happen in 2D semiconductor. In this simulation we only consider the ideal situation without the exciton. We only consider temperature making electron have more energy and the probability of forming electron-hole pair can be improved.

We suppose the kinetic energy of electron is $E_g$ when the temperature is $T$.

When the temperature goes up to $T + \Delta T$, the energy of electron is

$$E'_g = E_g + k_B \Delta T \tag{15}$$

$k_B$ is the Boltzmann constant and the energy gap $E'_g$ is

$$E'_g = V - E' = E_g - k_B \Delta T \tag{16}$$

Then the transmission probability $P$ can be expressed as:

$$P \sim \frac{(U/d)^{3/2} \cos\theta}{d_a^{1/2}} e^{-2\sqrt{2m(E_g - k_B \Delta T)}L/\hbar} \tag{17}$$

When the temperature is increasing, the state of electrons transport will be from the semiconducting phase to the metal phase. So the resistivity is decreasing with the improvement of temperature. But for the metal phase the resistivity is increasing with increasing temperature. Because the resistivity is dominated by the vibration of lattice or electron scattering in the metal phase. The metal phase is described by the Bloch individual-particle model. However, when the electrons transport from the metal phase to the superconducting phase, based on the Fermi-Dirac distribution, all the electrons nearly are under the Fermi surface and only few electrons near the Fermi surface can transport freely. The height of the barrier will decrease with temperature decreasing. So the transmission probability $P$ can be expressed as:

$$P \sim \frac{(U/d)^{3/2} \cos\theta}{d_a^{1/2}} e^{-2\sqrt{2m(E_g(T) - k_B T)}L/\hbar} \tag{18}$$

When $E_g(T) - k_B T \to 0$, it is the transition process from the metal phase to the superconducting phase. So the critical temperature $T_c$ is

$$T_c = E_g(T)/k_B \tag{19}$$

Now we consider a rectangular ordered lattice of $N \times M$ atoms. We introduce local structural disorder by assigning inconstant distance between the neighboring atoms, as shown in Fig. 4. The coordinate of each atom can be described as following:

$$X(m) = (m - 0.5)L \tag{20}$$

$$Y(n) = (n - 0.5)L \tag{21}$$

$L$ is the lattice constant. The vibration amplitude of every atom in the lattice can be described as following:

$$A_0 = C(T)L \tag{22}$$

$$A(n) = A_0 \sin(2\pi R) \tag{23}$$

So, the XY-coordinates of atoms in the lattice with local structural disorder are expressed as:

$$X(m) = (m - 0.5)L + C(T)L\sin(2\pi R) \tag{24}$$

$$Y(n) = (n - 0.5)L + C(T)L\sin(2\pi R) \tag{25}$$

$C(T)$ is a variable parameter, which is relate to the temperature, $R$ is the random number ($0<R<1$). When the temperature is improving, the vibration of crystal lattice becomes stronger. It leads to different lattice constants in the crystal. So the sine function $sin(2\pi R)$ is introduced and the value of $C(T)$ is increasing with temperature.

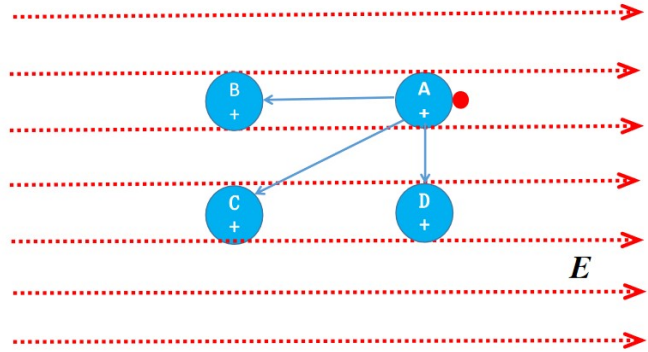

Fig. 5 The schematic of three representative processes of single-electron tunneling.

In the model we assume three representative processes of single-electron tunneling may occur in the single-layer semiconductor, as shown in the Fig. 5. When atomic core B, C and D are atomic core, atom A could transmit this electron to one of the three atomic cores. The tunneling rates of the three processes can be given from equation (14).

$$P_{A \to B} \sim \frac{(U/d)^{3/2} \cos\theta_{AB}}{d_A^{1/2}} e^{-2\sqrt{2mE_g}L_{AB}/\hbar} \tag{26}$$

$$P_{A \to C} \sim \frac{(U/d)^{3/2} \cos\theta_{AC}}{d_A^{1/2}} e^{-2\sqrt{2mE_g}L_{AC}/\hbar} \tag{27}$$

$$P_{A \to D} \sim \frac{(U/d)^{3/2} \cos\theta_{AD}}{d_A^{1/2}} e^{-2\sqrt{2mE_g}L_{AD}/\hbar} \tag{28}$$

In the crystal, we can get the follow relationship

$$\theta_{AB} < \theta_{AC} < \theta_{AD} \cong 90° \tag{29}$$

$$L_{AB} = L_{AD} < L_{AC} \tag{30}$$

So the relationship among the tunneling rates of these processes can be expressed as:

$$P_{A \to B} \gg P_{A \to C} \gg P_{A \to D} \tag{31}$$

Equation (31) implies that in a uniform array the charge moves only in 1D channels without any jumping of charge between adjacent channels. It means the most possible process from atom A to atom B may occur in the crystal (see Fig. 5). Next section we will study the FET and phototransistor. The underlying electron transport mechanism of FET and phototransistor and the emission mechanism of photoluminescence will be investigated.

4. **Results and discussion**

The electron transport properties of 2D semiconductor are determined by many interrelated parameters, such as charged impurities, defect and the production of exciton. In these parameters the production of exciton is most important for electron transport in 2D semiconductor. The exciton could make the field-effect mobility decrease at high temperature or high drain-source voltage.

In the simulation the probability of electron transporting from one covalent bond to an adjacent covalent bond with one hole is

$$P \sim \frac{(U/d)^{3/2} \cos\theta}{d_a^{1/2}} e^{-2\sqrt{2mE_g}L/\hbar} \tag{32}$$

IF $P \geq R$, $R$ is the random number $(0 \leq R \leq 1)$, then we consider this electron transport process can happen in the single-layer semiconductor.

4.1 *The characteristics of current-voltage in the single-layer FET*

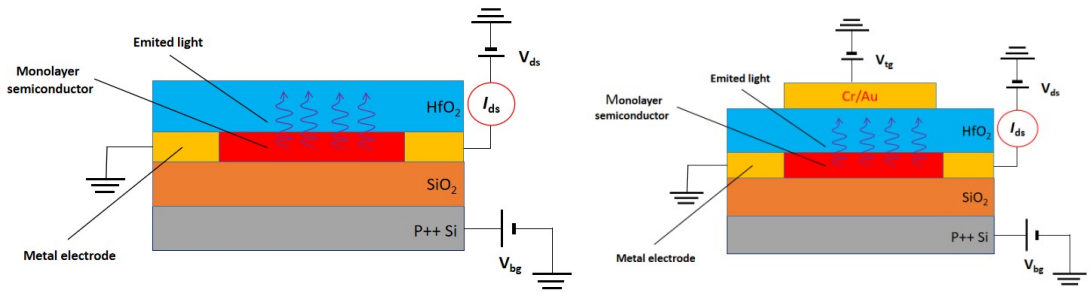

Fig. 6 Cross-sectional views of devices based on single-layer semiconductor in a single-gate (left) and dual-gate (right) configuration.

The devices are FETs in single-gate and dual-gate configurations shown in Fig. 6.

The gate electrode is used to control the concentration of carriers. When the voltage of back gate $V_{bg}$ is negative, due to electrostatic induction, the single-layer semiconductor will tend to be positive. So it will suppress the formation of electron-hole pairs. When the voltage of back gate $V_{bg}$ is positive, the single-layer semiconductor will tend to be negative. The electrostatic induction will motivate the formation of electron-hole pairs. Therefore a new control factor β will be introduced:

$$\beta = e^{C_1 V + C_2}/\left(e^{C_1 V + C_2} + 1\right) \tag{33}$$

$C_1$ and $C_2$ are constant, which are related to the dielectric material layer. When the gate voltage is very negative, $\beta \sim e^{C_1 V + C_2}$. The concentration of carriers is diminishes exponentially. When the gate voltage is very positive, $\beta \sim 1$, the density of carriers will approach saturation state and the current will reach the equilibrium state. In this equation we do not consider the emergence of exciton and the vibration of lattice. So the transmission probability $P$ can be expressed as:

$$P \sim \frac{(U/d)^{3/2}\cos\theta}{d_a^{1/2}} \beta_{bg}\beta_{tg} e^{-2\sqrt{2m(E_g - k_B \Delta T)}L/\hbar} \tag{34}$$

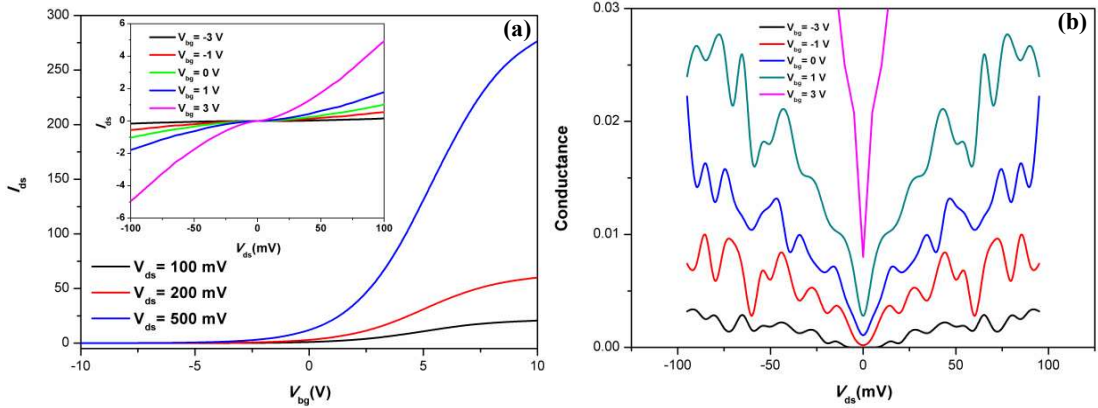

Fig. 7. Electrical characterization of the field-effect transistors based on monolayer semiconductor at room temperature. (a) $I_{ds}$-$V_{bg}$ curves acquired for different drain-source biases in the FET. Back-gate voltage is applied to the substrate. Inset: $I_{ds}$-$V_{ds}$ curves for $V_{bg}$ ranging from -3 to 3 V. (b) The differential conductance as a function of the back-gate voltage for the different values of $V_{bg}$= -3, -1, 0, 1, 3 V when $V_{ds}$ is 100 mV (from bottom to top).

In Fig. 7a are shown electrical characteristics of the transistor at room temperature. These gating characteristics are presented in Fig. 7a, showing a typical behavior of FETs with an n-type channel, which is very similar with the previous reports [15-24]. Such n-type doping is not coming from charge impurities. The difference between my simulation and the result of experiment is that we can not consider the production of exciton and the occurrence of electroluminescence. From the transmission probability, we can get this conclusion that the intensity of emitted light get maximum near the cathode. So the electroluminescence is localized in the region of the contacts, which is consistent with the result of experiment [25]. So in the experiment some energy is transferred to the formation of exciton and the recombination of electron-hole pairs causes electrically excited luminescence. So it is difficult to get the saturation state of current in the experiment. Nonlinear dependence of drain-source current $I_{ds}$ on bias voltage $V_{ds}$ clearly indicates the non-ohmic character, which is in agreement with the previous reports [15, 19, 23, 27, 28]. It is caused by the attraction force between electron and hole and we exclude the possibility that the field-effect behaviour is dominated by Schottky barriers at the contacts. Fig. 7b shows the differential conductance is increasing with the improving $V_{bg}$. It is the transition process from the suppression of formation of electron-hole pairs to the motivation of the formation of electron-hole pairs.

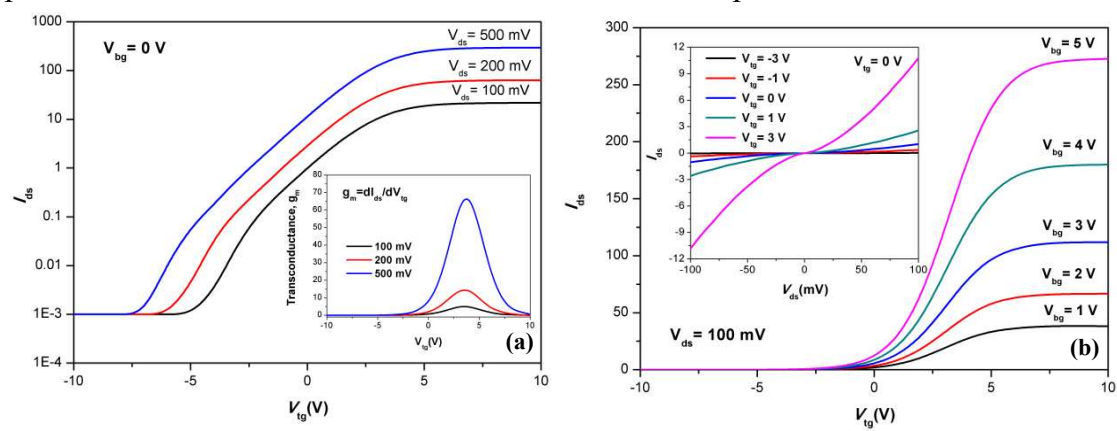

Fig. 8. Dual-gate control of the monolayer transistor. (a) $I_{ds}$–$V_{tg}$ curve for a bias voltage ranging from 100 mV to 500 mV at room temperature with the back gate grounded. Inset: transconductance $g_m =dI_{ds}/dV_{tg}$. (b) $I_{ds}$-$V_{tg}$ for values of $V_{bg}$ =1, 2, 3, 4 and 5 V. Inset: $I_{ds}$-$V_{ds}$ characteristics for different top-gate voltages $V_{tg}$ for drain voltages $V_{ds}$ reaching 3 V.

The corresponding transfer characteristic of dual-gate transistor is shown in Fig. 8a. For a bias of 100 mV, we observe an on-current of 10, current on/off ratio $I_{on}/I_{off}$ is $10^4$ for the ±5 V range of $V_{tg}$, an off-state current that is smaller than $10^{-3}$. The observed current variation for different values of $V_{tg}$ indicates that the field-effect behaviour of transistor is dominated by the single-layer semiconductor channel and not the contacts. The phenomenon of drain current saturation in the single-layer FET is appearing with the drain−gate conductance (transconductance $g_m=dI_{ds}/dV_{tg}$) close to zero (see the inset of Fig. 8a) at some point of the bottom-gate voltage. Fig. 8b shows the drain current is increasing when the improvement of the back-gate voltage and it finally reaches saturation current at some point of the top-gate voltage. The large degree of current control in the single-layer semiconductor is also clearly illustrated in the inset of Fig. 8b, where we plot the drain-source current versus drain-source bias for different values of voltage applied to the local gate. These results of simulation is consistent with the previous results of experiment [2, 15, 19].

## 4.2 *The effect of temperature on the characteristics of current-voltage in the single-layer FET*

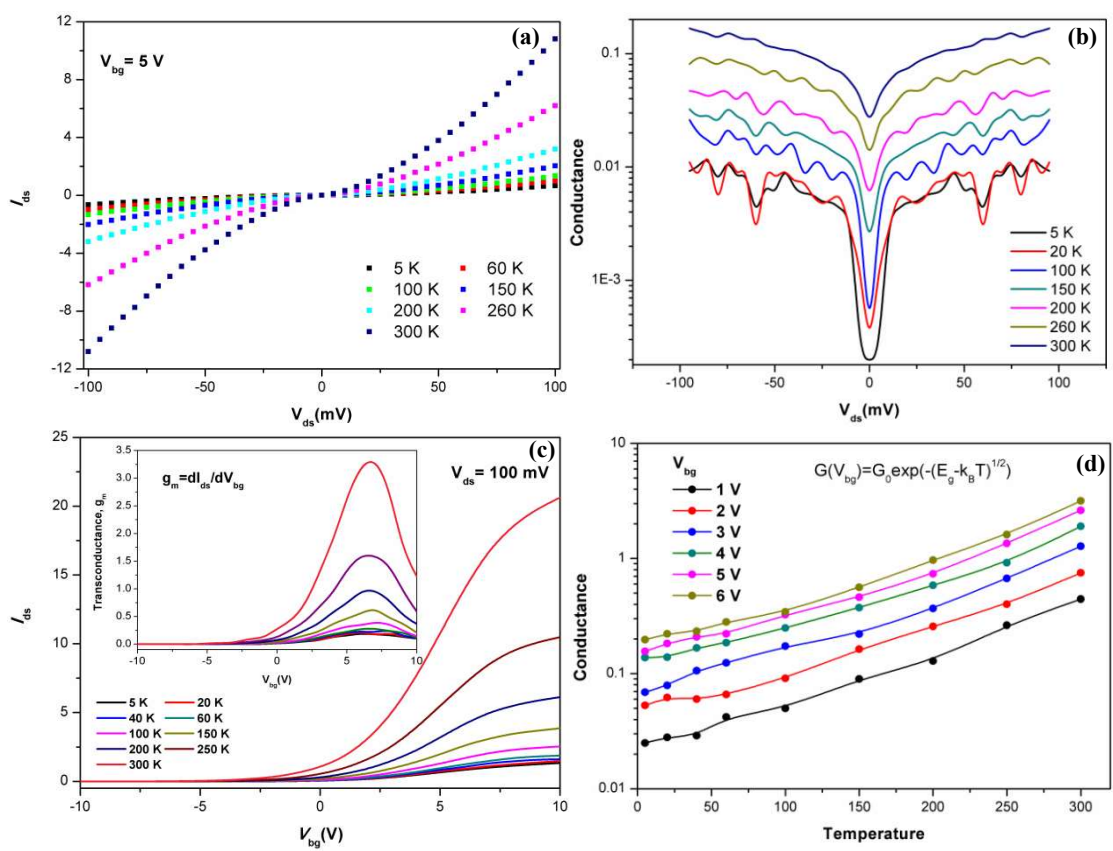

Fig. 9. Electron transport in the single-gate monolayer transistor. (a) The $I_{ds}$–$V_{ds}$ characteristics for several temperatures at $V_{bg}$ = 5 V. (b) The evolution of logarithmic plot of the differential conductance with temperature derived from $I_{ds}$-$V_{bg}$ characteristics shown in (a). (c) Current as a function of $V_{bg}$ for a single-gate monolayer transistor acquired at different temperatures at $V_{ds}$ =100 mV. The inset shows transconductance $g_m$ = d$I_{ds}$ / d$V_{bg}$ derived from $I_{ds}$-$V_{bg}$ characteristics shown in (c). (d) The plot of conductance for different values of $V_{bg}$. Solid lines are almost linear for limited regions of temperature and $V_{bg}$.

Fig. 9a shows the $I_{ds}$-$V_{ds}$ characteristics for several temperatures at $V_{bg}$ = 5 V for different temperatures and the nonlinear behaviours of curves are present for temperatures, which is caused by the tunneling barrier between electron and hole . it influences greatly the mobility movement. For low temperatures, there is no electron transport taking place up to a certain voltage that delivers the energy needed to overcome the covalent bond. This so-called threshold voltage decreases with increasing temperature, which is very well visible in the logarithmic plot of the respective differential conductances in Fig. 9b. Even at room temperature, the characteristic is still nonlinear. The current that is depended on the gate voltage for a single-gate device, reached the back-gate voltage $V_{bg}$ = 10 V, which is shown in Fig. 9c. Based on the transmission probability, electrons can get enough energy to overcome the constraint of covalent bond and have larger probability to get into the neighboring bond with the improvement of temperature. The transconductance $g_m = dI_{ds}/dV_{bg}$ is firstly increasing and then decreasing with the improvement of the back-gate voltage, which is shown in the inset of Fig. 9c. It is because the drain current approaches saturation state. We find that the temperature variation of conductance $G$ in a single-gate monolayer device (Fig. 9d), in the temperature regime, can be modeled with tunneling transport:

$$G = G_0(V_{bg})e^{-(E_g - k_B T)^{1/2}} \qquad (35)$$

where $E_g$ is the height of the barrier, $k_B$ is the Boltzmann constant and $G_0(V_{bg})$ is the voltage-dependent parameter. This model is different from the traditional thermally activated transport model:

$$G = G_0(T)e^{-E_a/k_B T}$$

Where $E_a$ is the activation energy, $k_B$ is the Boltzmann constant and $G_0(T)$ is the temperature-dependent parameter extracted from the fitting curves [19].

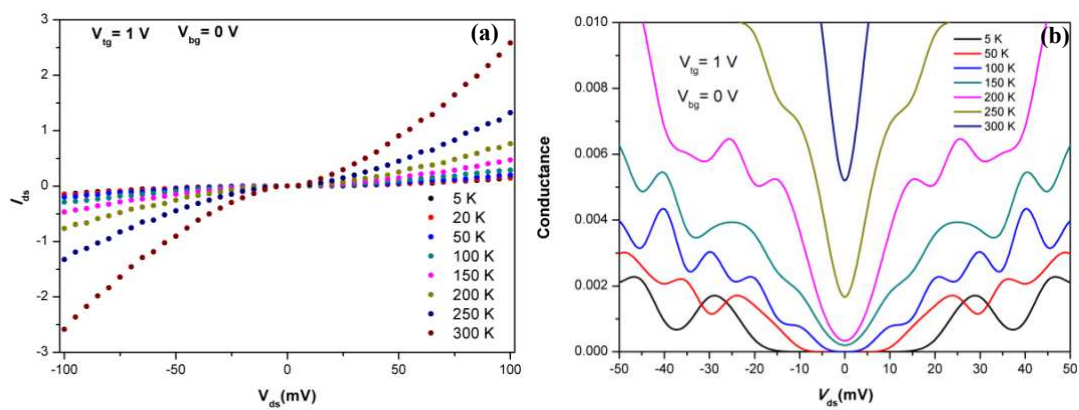

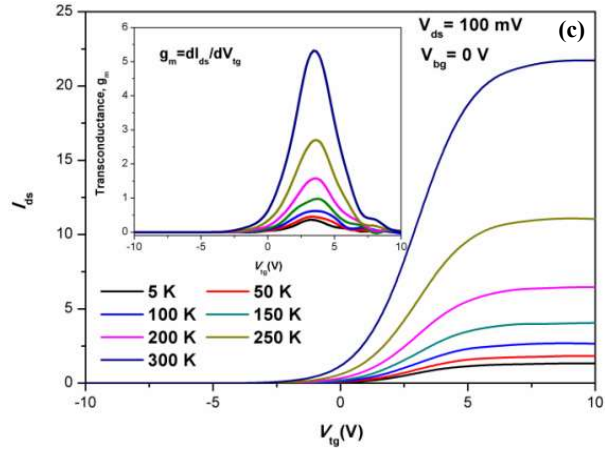

Fig. 10. Electron transport in the dual-gate monolayer transistor. (a) The $I_{ds}$–$V_{ds}$ characteristics for several temperatures at $V_{tg}$ = 1 V and $V_{bg}$ = 0 V. (b) The evolution of the differential conductance with temperature derived from $I_{ds}$–$V_{tg}$ characteristics shown in (a). (c) Current as a function of $V_{tg}$ for a dual-gate monolayer transistor acquired at different temperatures at $V_{ds}$ = 100 mV and $V_{bg}$ = 0 V. The inset shows transconductance $g_m$ = $dI_{ds}$ / $dV_{tg}$ derived from $I_{ds}$–$V_{tg}$ characteristics shown in (c).

The large degree of current control in the dual-gate transistor is also clearly illustrated in Fig. 10a, where we plot the drain–source current $I_{ds}$ versus drain–source bias $V_{ds}$ for different temperature at $V_{tg}$ = 1 V and $V_{bg}$ = 0 V. For low temperatures, the process of electron transport cannot takes place up to a certain voltage that delivers the energy needed to overcome the covalent bond. The threshold voltage decreases with increasing temperature, which is very well visible in the differential conductances in Fig. 10b. From the top-gating $I_{ds}$–$V_{tg}$ characteristics shown in Fig. 10c, the transconductance $g_m$ = $dI_{ds}/dV_{tg}$ can be derived. In the inset of Fig. 10c, we plot the 2D semiconductor device transconductance for different temperatures.

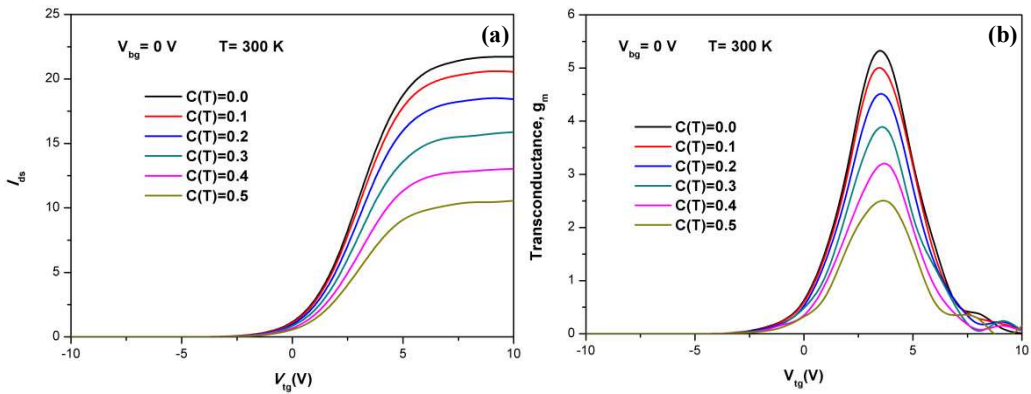

Fig. 11. Electron transport in the dual-gate monolayer transistor with different lattice vibrations at 300 K. (a) The $I_{ds}$–$V_{tg}$ characteristics for different lattice vibrations at $V_{bg}$ = 0 V. (b) The evolution of the differential conductance with lattice vibration derived from $I_{ds}$–$V_{tg}$ characteristics shown in (a).

There was the critical point of the metal–insulator transition in the single-layer semiconductor in some experimental reports [19, 24]. For low values of $V_{tg}$ or $V_{bg}$, the conductance decreases with temperature. Above some value of $V_{tg}$ or $V_{bg}$, the monolayer semiconductor enters a metallic state, with the conductance increasing as temperature is decreased. This transition point is the direct consequence of quantum interference effects of weak and strong localization. At lower carrier concentrations the system is in the insulating state and strong localization prevails. As the top-gate bias or temperature is increased, the system is driven into a metallic phase and weak localization seems to be the dominant effect. The observed quantum critical point of the metal–insulator transition in the 2D transistor is the consequence of a strongly correlated 2D electron gas. But in my opinion, this phenomena is caused by the formation of electron-hole pairs and the lattice vibration. When the temperature goes up, lots of exciton will be produced in the monolayer semiconductor and the vibration of atom will be strengthening. The exciton will be in the majority in some gate voltage. Fig. 11 shows that current and transconductance will decrease with the amplitude increasing. So in this case the current will decrease when the improvement of temperature. So the critical point of the metal–insulator transition will appear. The appearance of the metallic state should be in the conduction of high temperature or high gate positive voltage. In a word, the condition must be satisfy the formation of exciton and the large vibration of crystal lattice.

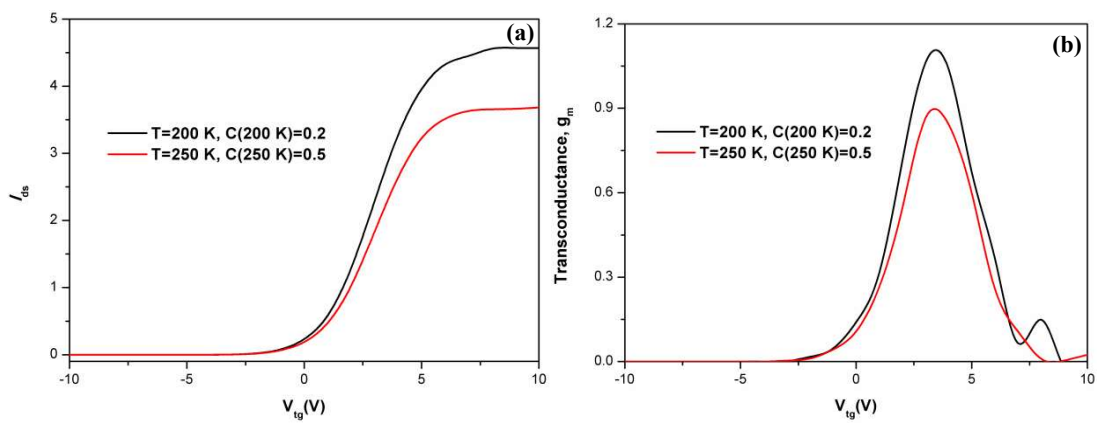

Fig. 12. Electron transport in the dual-gate monolayer transistor with different lattice vibrations at different temperatures. (a) The $I_{ds}$–$V_{tg}$ characteristics for different lattice vibrations at $V_{bg}$ = 0 V. (b) The evolution of the differential conductance with lattice vibration derived from $I_{ds}$–$V_{tg}$ characteristics shown in (a).

The transmission probability $P$ can be expressed as:

$$P \sim \frac{(U/d)^{3/2} \cos\theta}{d_a^{1/2}} \beta_{bg} e^{-2\sqrt{2m(E_g - k_B \Delta T)}L/\hbar} \quad (36)$$

$$A = \sqrt{2m(E_g - k_B T)}L \quad (37)$$

$$L \sim C(T) \tag{38}$$

$$A^2 \sim 2m(E_g - k_B T)C^2(T) \tag{39}$$

When the temperature is improving, the dislocation C(*T*) is increasing. So the value of $A^2$ may be increasing. Therefore the transmission probability could drop at certain temperature. Fig. 12 shows the $I_{ds}$-$V_{tg}$ characteristics for different lattice vibrations at different temperatures. We suppose the displacement C(*T*) is 0.5 when the temperature is 250 K and C(*T*) is 0.2 at 200 K. From Fig. 12, we can get the conclusion that the current and the conductance decrease when increasing the temperature. This is the mechanism of electron transport in the metal unless the atomic mass of some metallic elements are very heavy. So it is the transition mechanism from the insulator phase to the metallic phase.

### 4.3 *The characteristics of the single-layer semiconductor phototransistor*

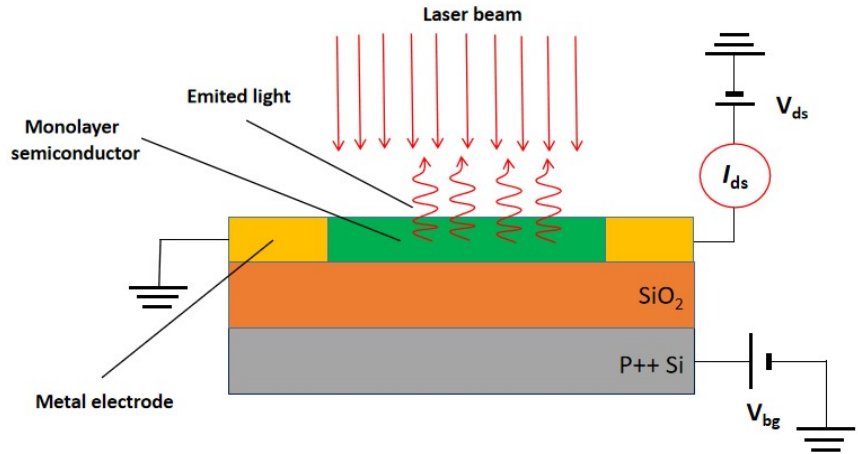

Fig. 13. The schematic view of the ultrasensitive single-layer semiconductor phototransistor.

In this section, we will investigate the single-layer based phototransistor and then study its light-induced electric properties in detail. Fig. 13 shows the schematic of the single-layer semiconductor phototransistor. Now we will investigate the mechanism of photonelectronin in the micro-atomic level by our theory, which is different from the traditional energy band theory. When the light exposure at the surface of phototransistor, some light is absorbed and the other light is reflected. When some photons are absorbed by the electron of covalent bond. There are three types of photo-electron in the phototransistor (see Fig. 14).

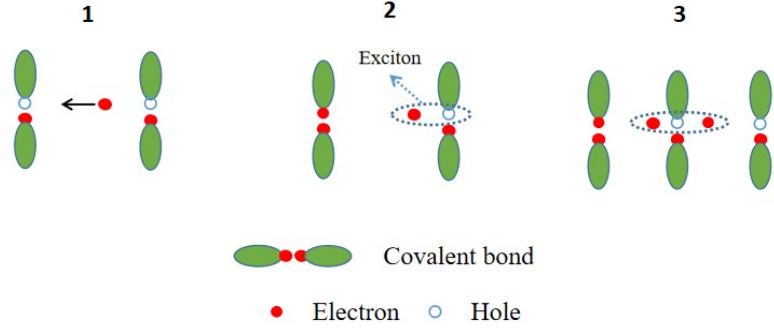

Fig. 14. Three emission processes of photoelectron in the phototransistor.

The first process (see the left of Fig. 14): photon-electron will recombine with the hole when the neighboring covalent bond has one hole with photoluminescence or return the old covalent bond with photoluminescence (see the left of Fig. 14). In this process photonelectron will have specific energy to get out of the Fermi surface and then the kinetic energy is almost zero. Due to the attraction of hole, this electron will go to the next covalent bond with one hole or return to the old covalent bond. When the absorbed energy of electron is more or less than the specific energy, photoelectron will not recombine with the hole. This specific energy means so-called band gap. The second process (see the middle of Fig. 14): Photon-electron and hole form the exciton when the neighboring covalent bond has no hole or recombine with photoluminescence. The third process (see the right of Fig. 14): two photon-electrons and one hole have chance to form the exciton with negative. In these processes only the first process has contribution to photoncurrent. So some photogenerated photon-electrons cannot be converted to the photoncurrent when the applied drain voltage is low. This is because a larger drain voltage can better drive photogenerated electrons to electrode, or suppress photogenerated electrons from the recombination. When the light power is increasing, the probability of formation of electron-hole pair will become larger per unit time. So the transmission probability $P$ can be expressed as:

$$P \sim \frac{(U/d)^{3/2}\cos\theta}{d_a^{1/2}} \beta_{bg} e^{-2\sqrt{2m(E_g - k_B \Delta T - nh\nu)}L/\hbar} \qquad (40)$$

where $n$ is the number of photons absorbed per unit time, which is proportional to the light power. $h$ is Planck constant, $\nu$ is the frequency of light.

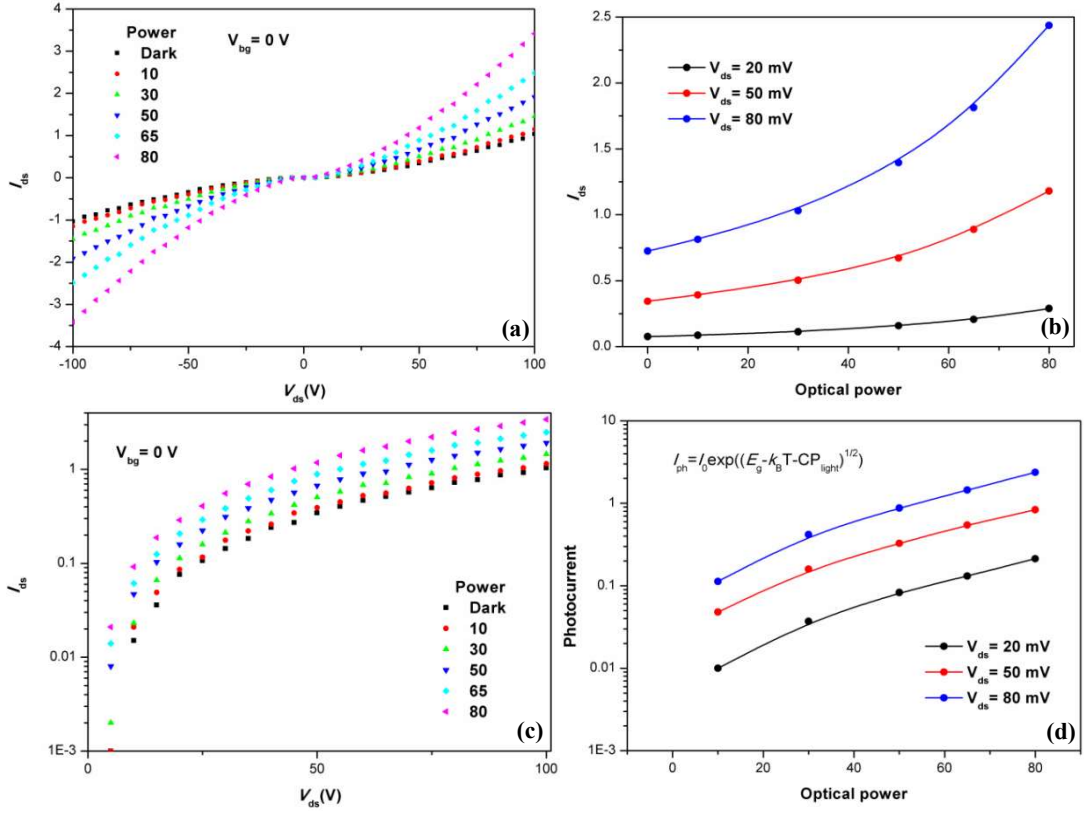

Fig. 15. (a) Electrical transport in single-layer phototransistor in the dark and under different at different illuminating optical power (10 to 80 ) at $V_g$ = 0 V. (b) Dependence of the drain current on optical power at different $V_{ds}$ (20, 50, and 80 mV). The nonlinear curves are fitting results. (c) The evolution of logarithmic plot of the drain current with optical power. (d) The photocurrent $I_{ph}$ in the different optical power.

Under the laser, the illumination on the single-layer phototransistor can generate an obvious photocurrent, which is power dependent. As shown in Fig. 15a and 15c, when the optical power increases from 10 to 80, the photocurrent increases gradually, even the applied drain voltage is 100 mV. To study the relationship between the output photocurrent ($I_{ph}$) and the incident optical power ($P_{light}$), the plot of $I_{ds}$ as function of $P_{light}$ is shown in Fig. 15b, based on the results in Fig. 15a. The $I_{ph}$ ($I_{ph} = I_{light,total} - I_{dark}$) under the constant drain voltage is nonlinearly proportional to $P_{light}$, which is shown in Fig. 15d. Such a relation has been described by the equation, $I_{ph} = I_0 \exp((E_g - k_B T - CP_{light})^{1/2})$, where $E_g$ is the height of the barrier, $k_B$ is the Boltzmann constant, $T$ is the temperature, $C$ is the fitting parameter, and $P_{light}$ is the incident optical power. The curves is consistent with the previous experimental reports [17, 26, 27].

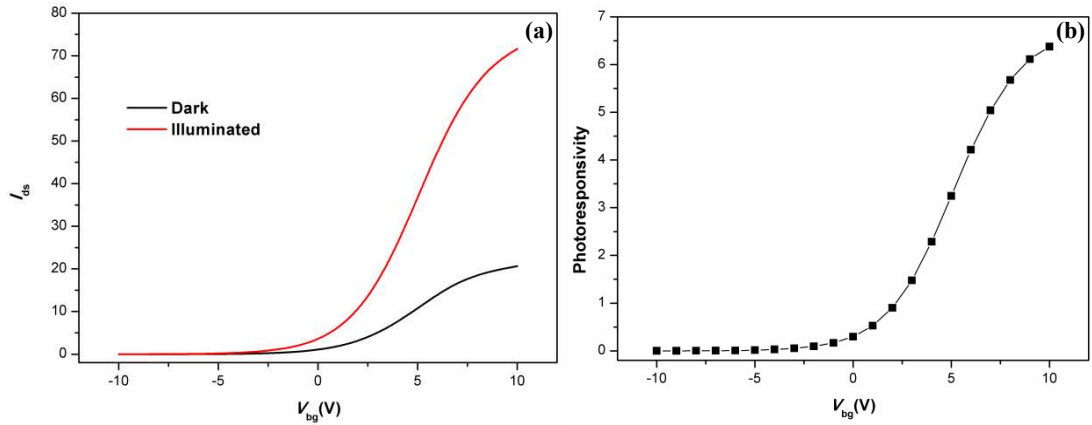

Fig. 16. (a) Typical output characteristics of single-layer phototransistor at gate voltage varied from -10 to 10 V ($P_{light}$ = 80). (b) Dependence of photoresponsivity on the gate voltage ($V_{ds}$ = 100 mV, $P_{light}$ = 80).

The photocurrent generation efficiency can be further enhanced by introduction of the gate voltage. In a typical experiment, if the gate voltage varies from -10 to 10 V, while $P_{light}$ and $V_{ds}$ are kept at 80 and 100 mV, respectively, the drain current under illumination is higher than that under dark (Fig. 16a). Moreover, the prompt photocurrent ON/OFF switching behavior is still well maintained in the range of applied gate voltage, which could be seen from the similar prompt photocurrent switching results under the different gate voltage as shown in Fig. 16b.

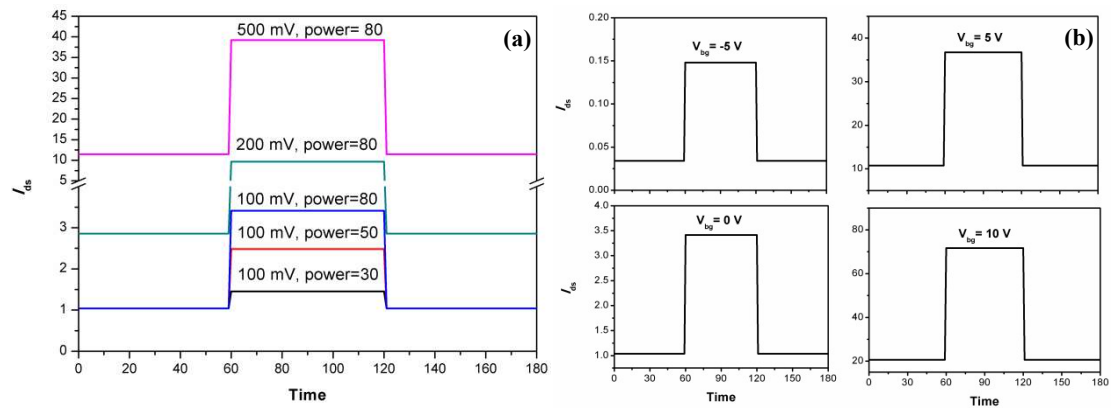

Fig. 17. (a) Photoswitching characteristics of single-layer phototransistor at different optical power ($P_{light}$) and drain voltage ($V_{ds}$). (b) Similar prompt photoswitching behavior at different gate voltage ($V_{ds}$ = 100 mV, $P_{light}$ = 80).

The photoswitching characteristic and stability of single-layer phototransistors were investigated at room temperature. In this simulation we suppose the switching duration for the current rise (from OFF to ON) or decay (from ON to OFF) process can be finished in unit time. As shown in Fig. 17a, the photocurrent as a function of time was measured under the alternative dark and illumination conditions at different optical power and drain voltage. The prompt photocurrent ON/OFF switching behavior is still well maintained in the range of applied gate voltage, which could be seen from the similar prompt photocurrent switching results under the different gate voltage as shown in Fig. 17b.

5. **Conclusion**

In this paper, a new theory has been developed to simulate the electron transport process in the single-layer semiconductor. In this process, single-gate FET, dual-gate FET and phototransistor are investigated by this theory in this paper. The effects of the back-gate voltage, top-gate voltage and temperature on the FET are studied by this new theory. The Coulomb attraction effect and single-electron tunneling are introduced in these simulations. The difference among the conductor, semiconductor and insulator are explained by this theory. A new control factor is decided by the back-gate voltage or the top-gate voltage is introduced in the transmission probability. In these process, the exciton would appear and the electroluminescence would happen. The results of simulation show the characteristics of the $I_{ds}$-$V_{ds}$ is symmetric and nonlinear in the single-layer semiconductor. However, the $I_{ds}$-$V_{bg}$ or $I_{ds}$-$V_{tg}$ is asymmetric and nonlinear. The back-gate voltage or the top-gate voltage that likes the switch can be used to control the drain current. The observation of a metal-insulator transition is explained in the single-layer semiconductor. The drain current saturation in single-layer FET, with the drain-source conductance close to zero are shown in the simulation. For phototransistor, the illumination on the single-layer semiconductor phototransistor can generate an obvious photocurrent. The photocurrent generation solely depends on the illuminating optical power at a constant drain or gate voltage. The prompt photoswitching behavior, controllable by the incident light, exhibits stable characteristics and the photoluminescence will happen in this process.

**Declaration of competing interest**

The authors declare that they have no known competing financial interests or personal relationships that could have appeared to influence the work reported in this paper.

**Acknowledgements**

This research was funded by the National Natural Science Foundation of China (Grant

Nos. 11627806 and U1909214), the National Key R&D Program of China (Grant No. 2016YFA0201002).

## Author contributions

Lianhua Zhang developed the ideas and theory for this work. Lianhua Zhang and the other authors prepared and edited the manuscript.